\pdfoutput=1
\DocumentMetadata{}
\documentclass[sigconf,nonacm]{acmart}

\usepackage{subcaption}
\usepackage[frozencache=true,cachedir=minted-cache]{minted}
\usepackage{multicol}

\definecolor{purple}{HTML}{DFDBEC}
\definecolor{orange}{HTML}{FAE2C3}
\setcopyright{none}
\copyrightyear{2024}
\acmYear{2024}
\acmDOI{XXXXXXX.XXXXXXX}

\acmConference[CIKM '24]{Make sure to enter the correct
  conference title from your rights confirmation email}{October 21--25,
  2018}{Boise, ID}
\acmPrice{15.00}
\acmISBN{978-1-4503-XXXX-X/18/06}

\begin{document}

\title{Large Language Models for JSON Schema Discovery}

\author{Michael J. Mior}
\email{mmior@mail.rit.edu}
\orcid{0000-0002-4057-8726}
\affiliation{%
  \institution{Rochester Institute of Technology}
  \streetaddress{102 Lomb Memorial Drive}
  \city{Rochester}
  \state{New York}
  \country{USA}
  \postcode{14612-5608}
}

\renewcommand{\shortauthors}{Mior}

\begin{abstract}
Semi-structured data formats such as JSON have proved to be useful data models for applications that require flexibility in the format of data stored.
However, JSON data often come without the schemas that are typically available with relational data.
This has resulted in a number of tools for discovering schemas from a collection of data.
Although such tools can be useful, existing approaches focus on the syntax of documents and ignore semantic information.

In this work, we explore the automatic addition of meaningful semantic information to discovered schemas similar to information that is added by human schema authors.
We leverage large language models and a corpus of manually authored JSON Schema documents to generate natural language descriptions of schema elements, meaningful names for reusable definitions, and identify which discovered properties are most useful and which can be considered ``noise''.
Our approach performs well on existing metrics for text generation that have been previously shown to correlate well with human judgement.
\end{abstract}




\maketitle

\section{Introduction}

Large language models (LLMs) such as ChatGPT have had a significant impact in all fields of computer science.
One of the most common applications is natural language question answering.
However, there have been many applications of LLMs for software development.
Models such as StarCoder~\cite{StarCoder}, InCoder~\cite{InCoder}, and OpenAI's codex have been trained on large corpora of source code and can be used for intelligent autocompletion in existing software applications or even to develop new applications from scratch.
Such models manage to learn useful representations of source code that can be used for other tasks.

In this work, we explore several applications of LLMs in the domain of schema discovery for semi-structured data.
Such schemas are typically represented in text format and are often embedded in program source code.
As such, we expect many such schemas to be included in the training corpora of LLMs for code.
We can take advantage of these pre-trained models to improve the quality of the schemas discovered from collections of JSON documents using existing schema discovery tools.

First, we use properties discovered from datasets to provide natural language descriptions of elements of a schema.
Second, we demonstrate how definitions of repeated subschemas extracted from a schema can be given semantically meaningful names.
Finally, we build a classifier to identify properties of schemas that are included in manually authored schemas to better align automatically discovered schemas with those that are handcrafted.
All of our results are validated using a dataset of several hundred real-world JSON Schemas.

Consider the example schemas shown in Figure~\ref{fig:example_schemas}.
The schema on the left was manually authored.
It has been slightly modified from the original version for illustrative purposes, but exhibits similar patterns to real-world schemas.
Note that the original schema in Figure~\ref{fig:orig_schema} contains natural language descriptions of several schema elements.
There is also a definition called \texttt{WebJob} that is appropriately named to match the semantics of the underlying data.

In contrast, the schema in Figure~\ref{fig:discovered_schema} is typical of what might be created by a tool for JSON schema discovery based on a collection of documents.
In this case, the schema contains no descriptive text and the definition is given the unhelpful name of \texttt{defn0} since no attempt is made to assign any semantic meaning.
Furthermore, there are a number of unhelpful properties that are defined within the schema.
For example, the property \texttt{filePath} is assigned a maximum and minimum length of 19 and 112, respectively, since these were the lengths that were observed in the documents used for discovery.
It is unlikely that these values represent any meaningful constraint on the underlying domain, which is a likely reason that these properties were not included in the manually authored schema.

\begin{figure*}
    \centering
    \begin{subcaptionblock}{0.43\textwidth}
{\scriptsize
    \begin{minted}[escapeinside=||]{json}
{
  "definitions": {"|\colorbox{purple}{WebJob}|": {
    "type": "object", "additionalProperties": false,
    "properties": {"filePath": {"type": "string"}}}}},
  }},
  "properties": {
    "WebJobs": {
:     "description": |\colorbox{purple}{A list of Azure Webjobs}|,
      "type": "array", "items": {"$ref": "#/definitions/WebJob"}}},
  "required": ["WebJobs"],
  "title": |\colorbox{purple}{JSON Schema for Azure Webjobs collection files}|,
  "type": "object"
}
    \end{minted}
}
\caption{Original schema}\label{fig:orig_schema}
    \end{subcaptionblock}
    \begin{subcaptionblock}{0.5\textwidth}
{\scriptsize
    \begin{minted}[escapeinside=||]{json}
{
  "definitions": {"|\colorbox{orange}{defn0}|": {
     "properties" : {
       "filePath" : {"type" : "string", |\colorbox{orange}{"minLength" : 19}|, |\colorbox{orange}{"maxLength" : 112}|}}},
  }},
  "type" : "object", "additionalProperties" : false,
  "properties" : {
    "WebJobs" : {
      "type" : "array",
      "items" : {
        "$ref": "#/definitions/defn0",
        "type" : "object", "additionalProperties" : false}
      |\colorbox{orange}{"minItems" : 1}|, |\colorbox{orange}{"maxItems" : 12}|}},
  "required" : [ "$schema", "WebJobs" ]
}
\end{minted}
}
    \caption{Discovered schema}\label{fig:discovered_schema}
    \end{subcaptionblock}
    \caption{Azure Webjob schema example}\label{fig:example_schemas}
    \Description{Two JSON Schemas describing Azure AzureWebJobs. The first contains a definition with a menaingful name and natural language descriptions. The second schema has an automatically generated name, no descriptions, and some irrelevant properties.}
\end{figure*}

\section{Background}\label{sec:background}

Our work aims to leverage two distinct areas of research to improve the management of schemas for JSON documents.
Existing work on schema discovery provides schemas that are correct, but lack several useful common elements that are present in schemas written by human authors.
Our goal in this work is to combine the scalability and ease of use provided by automatic schema discovery while retaining the benefits of manually authored schemas by leveraging LLMs for code.
Below we describe the existing work we leverage from schema discovery and LLMs.

\subsection{Schema Discovery}\label{subsec:discovery}

Unlike relational databases, collections of documents in JSON format often do not come with an associated schema.
This complicates the further analysis of the data, as consumers are often forced to use a ``guess and check'' approach to developing data pipelines.
Analysts inspect a few documents and craft the assumptions necessary to construct their analysis and then execute the pipeline on the entire data set to validate these assumptions.
This process is slow and error-prone.
To relieve this burden, many authors have proposed \emph{discovering} a schema from a collection of documents.
Baazizi et al.~\cite{BaaziziCGS17,BaaziziBCGS20,BaaziziCGS19,BaaziziLCGS17} present an approach based on discovering a schema for individual JSON documents and then merging each of these schemas to create a final schema for an entire collection.
Such schemas have proved useful for managing schemas for NoSQL databases as used in the Josch~\cite{Josch} schema management tool.

\subsection{Large Language Models}

Large language models have shown significant potential for use in a wide variety of data integration tasks~\cite{FoundationModels,Fernandez2023}.
In our case, it is particularly relevant that many of these models have been trained on large corpora of open source code~\cite{InCoder,StarCoder,CodeBERT}.
Such models are commonly used for tasks such as code completion and code understanding and summarization.

\section{Schema Annotation}

While past work on schema discovery has focused primarily on recovering descriptive structural information, our goal in this work is to add meaningful semantic information to schemas.
First, by generating meaningful descriptions for schema elements and second, by providing meaningful names from any repeated structures that have been identified.
We also use the model to identify which keywords are actually expected to express meaningful constraints on the underling domain.
In the following sections, we describe the construction of the training data for each of our tasks, as well as the prompts used to fine-tune the LLM.



\subsection{Natural Language Descriptions}


A well-written manual schema involves not only structural information, but also informative descriptions of various schema elements.
Typical schema discovery mechanisms focus solely on the structure and do not address this additional semantic information.
In fact, in JSON Schema Store~\cite{SchemaStore}, an open source repository of JSON Schemas, we find over 90\% of schemas make use of natural language descriptions.
This gives us a rich source of training data since we have existing schema elements with descriptions attached.

Note that we consider not just descriptions for the entire schema, but also individual properties within the schema.
Our dataset of descriptions extracted from JSON Schema Store includes over 50,000 schema fragments that have an associated description.
The goal of our model in this case is to produce a natural language description for each schema element.
In this case, we prompt the LLM with \texttt{Generate a short description for the given JSON Schema}.
The first 100 characters of schema fragment is then provided to the model as input.
We take any output the model produces as the corresponding description that should be included in the schema.

\subsection{Structure Naming}

A common feature of JSON Schema is \emph{definitions} that identify repeated structures used throughout a document.
For example, a dataset that makes significant use of geographic data may have several positions identified using the keys \texttt{"lat"} and \texttt{"lon"}.
We can reduce repetition and make the schema more comprehensible by creating a definition of objects with these properties and including a reference to this object wherever these properties are used.
For a human developer, it might be easy to decide that this definition should be named \texttt{"point"} (or something similar).
This problem becomes much more challenging when automating schema discovery.
The process of finding such definitions can involve identifying repeated use of the same set of keys within an object.
Ideally, we would like any schema discovery process to be able to produce names for these definitions similar to the names that would be selected by a human familiar with the dataset.
In our example above with \texttt{lat} and \texttt{lon}, a suitable name for this definition may be \texttt{location}.
Rather than generating a meaningless name for each definition (e.g. \texttt{defn3}), our goal is to generate semantically meaningful names.

Across our example dataset, we have over 6,000 definitions we use to build our model.
Each training example given to the model includes the name of a definition as well as the schema represented by the definition.
For this task, we use the prompt \texttt{Generate a name for the given JSON Schema definition consisting of a single programming language identifier}.
This generated identifier is used as the name of the definition.

\subsection{Property Selection}

One of the challenges with schema discovery is generating properties that are actually relevant to developers who consume a schema.
For example, one property that a schema discovery tool may find is the minimum length of a string.
However, since the discovery of this property is based on sampling data values, assuming lengths are normally distributed, we are unlikely to observe the true minimum for smaller sample sizes.
The end result is that, although the property values discovered match the documents used during discovery, they may not be representative of the true constraints on the underlying domain.
The schemas discovered from datasets by the existing tools are, therefore, \emph{descriptive} of the dataset itself.
However, they are less useful for describing any constraints that should hold on new data items.

To create schemas that are more reflective of the underlying domain of the data, rather than the specific dataset used for discovery, our goal is to learn which properties and values are useful from manually authored schemas.
These manually authored schemas are generally written with the goal of validating data according to domain constraints and are not tied to any specific sample of data.
By learning from these schemas, we can adapt existing discovery tools to produce schemas that are closer to those that would be written by a human author.

\section{Methodology}

In order to create the annotations mentioned in the previous sections, we make use of a large corpus of real-world schemas from the JSON Schema Store~\cite{SchemaStore}, an open source catalog of JSON Schemas.
We collected a total of 657 valid schemas, most of which are configuration formats for software development tools.
From these schemas, we extracted examples of each annotation that we are attempting to apply.
These schemas have examples of both definitions and natural language descriptions, which we collect along with the schema fragment to which they apply with examples shown in Figure~\ref{fig:example_fragments}.

\begin{figure}
{\tiny
\begin{minted}{json}
{"type": "object", "properties": {"message-contain": {"type": "string"}}
{"type": "integer", "minimum": 0}
{"type": "array", "items": {"type": "string"}, "minItems": 1}
{"type": "string", "format": "date-time"}
{"type": "string", "enum": [null, "database", "objectStorage"]}
\end{minted}
}
    \caption{Example schema fragments for training}\label{fig:example_fragments}
    \Description{Small pieces of JSON Schemas extracted from full JSON Schema documents.}
\end{figure}

Collecting training data for property selection is more complicated since we need examples of both properties that should and should not be included in the resulting schema.
Generating positive examples of properties that should be included is straightforward by sampling properties in the example schema.
It is less obvious how to generate negative examples since, by definition, these are for properties that do not exist in our collection of schemas.
To generate negative examples, we also need a collection of schemas discovered from a data sample, meaning we also need example documents that conform to each schema.

As mentioned above, many of the schemas in the JSON Schema Store are for configuration files for various software tools.
These configuration files often have prescribed names that must be used so that the corresponding tool can find them.
Where such prescribed names exist, they are annotated in metadata associated with the schema in the JSON Schema Store.
We use these names to perform a search for open source code on GitHub to discover example documents with the corresponding filenames.
These documents are then validated according to each schema to ensure that the match in filename was not simply due to coincidence.
Finally, we extract a schema from each collection of documents using the JSONoid\cite{mior2023jsonoid} schema discovery tool.

To generate negative examples, we only consider properties that were used elsewhere in the schema.
This avoids the case where the schema author was not aware of a particular property or chose to avoid its use entirely for a particular schema.
Using the discovered schemas, we sample properties used somewhere in a schema, but not at the specific location in the schema, as shown in Figure~\ref{fig:property_selection}.

For the positive examples, the properties \texttt{format} and \texttt{minItems} were actually used in the manually authored schema.
In the first negative example, the property \texttt{minLength} was used somewhere in the schema, but not specifically for the property \texttt{adopt-info}.
This suggests that the author is aware of this property, but decided not to apply it in this particular setting.
The value \texttt{3} of the property observed here is instead obtained from the automatically discovered schema.
By learning to differentiate between these positive and negative examples, we can create schemas that are more similar to those created by human authors.

\begin{figure}
\raggedright
\textbf{Positive} \\\smallskip
{\tiny
\texttt{format}: \mintinline{json}{{"properties":{"uri":{"format":"uri"}}}} \\
\texttt{minItems}: \mintinline{json}{{"properties":{"Default fields":{"minItems":1}}}}
}

\smallskip

\textbf{Negative} \\\smallskip
{\tiny
\texttt{minItems}: \mintinline{json}{{"properties":{"adopt-info":{"minLength":3}}}}\newline
\texttt{uniqueItems}: \mintinline{json}{{"properties":{"classifications":{"uniqueItems":true}}}}
}

    \caption{Examples used for property selection}\label{fig:property_selection}
    \Description{Positive and negative examples of properties that should (and should not) be included in a JSON Schema.}
\end{figure}

\section{Evaluation}

In order to evaluate the performance of our model at generating useful schemas, we take the approach of removing the information we aim to generate from a test set of schemas and comparing whether the approach we use generates information closely matching the original schema.
We note, however, that since these schemas are manually authored, they do contain some information we would not expect to find in schemas output from a discovery tool.
Specifically, we remove any instances of keywords such as \texttt{title}, \texttt{description}, and \texttt{\$comment} anywhere in the schema.
These natural language explanations of schema elements would provide an unfair advantage to our discovery process.

For our base model, we use Code Llama~\cite{CodeLlama}, an open source foundation model for code.
We selected this model since JSON data is very similar (and often embedded in) programming language source code.
We performed fine tuning on the base Code Llama model using LoRA~\cite{hu2022lora} for a single epoch.
Other relevant hyperparameters are given in Table~\ref{tbl:hyperparams}.
We split our 
The fine tuning process completed in under 2 hours using a single NVIDIA A100 GPU.
We compare against the original Code Llama model without fine tuning as well as a simple baseline for each annotation.
All source code for our model is provided~\footnote{\url{https://anonymous.4open.science/r/json-schema-annotations-E3DC/}}.

\begin{table}
    {\small
    \begin{tabular}{c|c}
         Hyperparameter & Value  \\ \hline
         Batch size & 8 \\
         Learning rate  & (max=$2\times 10^{-4}$, min=$8\times 10^{-6}$, schedule=cosine)  \\
         Optimizer & AdamW ($\beta_1=0.9$, $\beta_2=0.95$) \\
         Weight decay & 0.016 \\
         LoRA dropout & 0.008 \\
         LoRA rank & 32 \\
         LoRA alpha & 16
    \end{tabular}
    }
    \caption{Model hyperparameters}\label{tbl:hyperparams}
\end{table}


\subsection{Natural Language Descriptions}

To assess our natural language descriptions, our primary metric is BERTScore~\cite{BERTScore}, a metric for text generation that has been shown to correlate positively with human judgement.
BERTScore uses BERT embeddings~\cite{BERT} to compare the candidate generated text (in our case, a proposed description) with a reference text (the original description in the schema.
Since our context is JSON Schema, which closely resembles source code, we use the CodeBERT~\cite{CodeBERT} model with BERTScore.
CodeBERT is a variant of BERT trained on both natural language and program source code, which is particularly relevant for our setting.
As an additional baseline, we also consider a pre-trained model for JavaScript code generation~\cite{CodeTrans} based on the T5~\cite{T5} architecture.
The scores for each model are shown in Table~\ref{tbl:eval_desc} along with the ROUGE-L~\cite{ROUGEL} and BLEU~\cite{BLEU} scores for comparison.
Our model performs the best on all metrics.
Figure~\ref{fig:example_desc} shows example descriptions generated by our model with both the highest and lowest scores.
We note that in the low-scoring case, the model has very limited context to work with in generating a description.
As future work, we plan to explore what additional context from the schema or data files might be helpful to improve the quality of our generation.

\begin{table}
    \begin{tabular}{c c c c}
     Metric & Ours & Code Llama & T5 \\\hline
     BERTScore & \textbf{0.763} & 0.707 & 0.729 \\
     ROUGE-L & \textbf{0.185} & 0.054 & 0.125 \\
     BLEU & \textbf{0.025} & 0.001 & 0.012\\
    \end{tabular}
    \caption{Evaluation metrics for natural language descriptions}\label{tbl:eval_desc}
    \vspace*{-2em}
\end{table}

\begin{figure}
\raggedright
{\scriptsize
\begin{minted}{json}
{"tsConfig":{"oneOf":[{"type":"string"},
                      {"type":"array","items":{"type":"string"}}]}}
\end{minted}
}
{\small\textbf{Original}: The name of the TypeScript configuration file.\\
\textbf{Ours}: TypeScript configuration file path.\\ 
\textbf{BERTScore}: 0.89}
\bigskip
{\scriptsize
\begin{minted}{json}
{"laquo_space":{"type":"string"}}
\end{minted}
}
{\small\textbf{Original}: Typographical symbol \\
\textbf{Ours}: Quotes are paired with a space in between. \\
\textbf{BERTScore}: 0.66}
\caption{Example descriptions generated by our model}\label{fig:example_desc}
\Description{An example of both a high and low-quality description generated by our proposed model.}
\end{figure}

\subsection{Definitions}

To evaluate the definition names generated by our model, we use similarity scores from VarCLR~\cite{VarCLR}, a semantic representation of variable names.
Definition names are similar to variable names in that they are identifiers used as a label for a unit of data.
VarCLR similarity scores have been shown to correlate well with human judgement of quality variable names.
Table~\ref{tbl:eval_defn} shows the scores for all the models we consider in our evaluation.
In this case, our model shows more than a 40\% increase over the next best model.
Examples of definition names and their associated scores are given in Figure~\ref{fig:example_defn}.
As in the case of descriptions above, for the low scoring case, our model can suffer from the limited context available which we plan to address in future work.

\begin{table}
    \begin{tabular}{c c c c}
     Metric & Ours & Code Llama & T5 \\\hline
     VarCLR & \textbf{0.517} & 0.335 & 0.369 \\
    \end{tabular}
    \caption{Evaluation metrics for definition naming}\label{tbl:eval_defn}
\end{table}

\begin{figure}
    \raggedright
    {\scriptsize
    \begin{minted}{json}
    {"enum":["proprietary", "0BSD", …], "type": "string"}
    \end{minted}
    }
    {\small\textbf{Original}: enums:spdx-license \\
    \textbf{Ours}: SpdxLicenseId\\ 
    \textbf{VarCLR}: 0.77}
    \bigskip
    {\scriptsize
    \begin{minted}{json}
    {"type": "string"}
    \end{minted}
    }
    \small{\textbf{Original}: umbracoUnattended \\
    \textbf{Ours}: UpdateModuleParams \\
    \textbf{BERTScore}: 0.21}
    \caption{Example definition names generated by our model}\label{fig:example_defn}
    \Description{An example of both a high and low-quality definition names generated by our proposed model.}
\end{figure}

\subsection{Property Selection}

Our evaluation metric in the case of property selection is simply the accuracy of the model in classifying whether a property should be selected.
Note that in this case we cannot evaluate against T5 since T5 is only able to perform code completion, which is not sufficient for this use case.
After fine tuning, our model achieves an accuracy of 90.5\% compared to 62.7\% for the base model.


\section{Related Work}

As discussed in Section~\ref{subsec:discovery}, there has been significant previous work on discovering schemas from collections of JSON documents.
However, previous work focused primarily on identifying the structure and type of data present in documents, without attempting to infer any semantic meaning.
Since in many cases human users are the primary consumers of JSON Schema, providing semantically meaningful information, which is the primary goal of this work, is incredibly useful.
Unfortunately, schemas discovered directly from data often have a structure different from those designed by human authors.
For example, some authors have identified cases where the resulting schema is ambiguous~\cite{namba2021enhancing,Jxplain}.
Others have identified cases where certain features of the JSON Schema language have been exploited to produce more semantically accurate schemas by capturing dependencies in the data~\cite{KlessingerFGKSS23,klessinger2023capturing}.

However, we are unaware of existing approaches that augment schema discovery by adding natural language elements to discovered schemas.
We note that our approaches here can be applied on top of any existing tool which discovers JSON Schema.
Thus, any existing schema discovery tool can benefit from the ability to add the additional semantic information that we describe here.

\section{Conclusions and Future Work}

Although the semantic information that we add to the schemas here is useful, we believe that additional tool support for leveraging these data could further increase the utility.
For example, our classifier identifies which discovered properties should be exposed in a schema to mimic manually written schemas.
However, it is possible that some users may want more or less detail when viewing a schema.
To this end, a possible additional use case for such a classifier could be as a control to adjust the level of detail.
Properties that appear to be less useful could be hidden to get an overview of the entire schema and then shown whenever the user has a desire to drill down deeper.
This could be implemented on top of our approach by examining the logits output from the model rather than only the final prediction.
Furthermore, we expect the performance of the model can be significantly improved by including relevant context from the entire schema.


\onecolumn
\begin{multicols}{2}
\bibliographystyle{ACM-Reference-Format}
\bibliography{references}
\end{multicols}


\end{document}